\documentclass[draft]{llncs}

\usepackage{mathtools}
\usepackage{makecell}
\usepackage{graphicx}
\usepackage{amsmath}
\usepackage{pdflscape}
\usepackage{rotating}

\usepackage{enumitem}

\usepackage{amsmath,amssymb}

\renewcommand{\figurename}{Fig.{}}

\usepackage{changepage}

\usepackage[utf8x]{inputenc}

\usepackage{textcomp,marvosym}

\usepackage{cite}

\usepackage{nameref,hyperref}

\usepackage[right]{lineno}

\usepackage{microtype}
\DisableLigatures[f]{encoding = *, family = * }

\usepackage[table]{xcolor}

\usepackage{todonotes}

\usepackage{array}

\usepackage{listings} 

\usepackage{subcaption}

\usepackage[british]{babel}
\usepackage{hhline}
\usepackage{multirow}
\usepackage[figurename=Fig]{caption}

\usepackage{tkz-orm}

\usepackage{verbatim}
\usepackage{pgfplots}

 \usepackage{tikz}
 \usetikzlibrary{arrows}
 \usepackage{xparse}
\usetikzlibrary{matrix,backgrounds}
\pgfdeclarelayer{myback}
\pgfsetlayers{myback,background,main}

\tikzset{mycolor/.style = {line width=1bp,color=#1}}%
\tikzset{myfillcolor/.style = {draw,fill=#1}}%

\NewDocumentCommand{\highlight}{O{blue!40} m m}{%
\draw[mycolor=#1] (#2.north west)rectangle (#3.south east);
}

\NewDocumentCommand{\fhighlight}{O{blue!40} m m}{%
\draw[myfillcolor=#1] (#2.north west)rectangle (#3.south east);
}
 \usetikzlibrary{matrix,decorations.pathreplacing, calc, positioning}

\newcolumntype{+}{!{\vrule width 2pt}}

\newlength\savedwidth

\usepackage{array}
\newcolumntype{L}[1]{>{\raggedright\let\newline\\\arraybackslash\hspace{0pt}}m{#1}}
\newcolumntype{C}[1]{>{\centering\let\newline\\\arraybackslash\hspace{0pt}}m{#1}}
\newcolumntype{R}[1]{>{\raggedleft\let\newline\\\arraybackslash\hspace{0pt}}m{#1}}

\newcommand{\booleans}{\mathbb{B}}

\newcommand{\maps}{\rightarrow}

\newcommand{\union}{{\cup} }

\newcommand{\powerset}[1]{2^{#1}}
\newcommand{\intersection}{{\cap} }
\newcommand{\intersect}{\intersection}

\newcommand{\limplies}{\Rightarrow}
\newcommand{\lxor}{\oplus}
\newcommand{\Land}{\bigwedge}
\newcommand{\Lor}{\bigvee}

\newcommand{\nodes}{N}
\newcommand{\mols}{M}
\newcommand{\nlabel}{L}
\newcommand{\edges}{E}
\newcommand{\pairs}{\mathcal{P}}
\newcommand{\nodef}{f}
\newcommand{\edgef}{g}

\newcommand{\zthree}{\textsc{Z3}}
\newcommand{\ourtool}{\textsc{VTSSynth}}
\newcommand{\depqbf}{\textsc{DepQBF}}

\newtheorem{df}{Definition}

\begin{document}

\title{Synthesis for Vesicle Traffic Systems}

\author{Ashutosh Gupta$^1$ \and Somya Mani$^2$ \and Ankit Shukla$^3$}
\institute{IIT Bombay \and IBS-CSLM, South Korea \and NCBS, Bangalore}


\date{\today}

\maketitle

\begin{abstract}
Vesicle Traffic Systems (VTSs) are the material transport 
mechanisms among the compartments inside the biological cells.
The compartments are viewed as nodes that are labeled with the
containing chemicals and the transport channels
are similarly viewed as labeled edges between the nodes.
Understanding VTSs is an ongoing area of research and for many cells
they are partially known.
For example, there may be undiscovered edges, nodes, or their labels
in a VTS of a cell.
It has been speculated that there are properties that the VTSs must
satisfy.
For example, stability, i.e., every chemical that is leaving a
compartment comes back.
Many synthesis questions may arise in this scenario, where we want
to complete a partially known VTS under a given property.
In the paper, we present novel encodings of the above questions
into the QBF (quantified Boolean formula) satisfiability problems.
We have implemented the encodings in a highly configurable tool
and applied to a couple of found-in-nature VTSs and 
several synthetic graphs.
Our results demonstrate that our method can scale up to the graphs of interest.
%
%


\end{abstract}

\section{Introduction}
\label{sec:intro}

Eukaryotic cells, including human cells, consist of multiple membrane-bound compartments.
Material is transported among these compartments by the vesicle
transport system (VTS).
Briefly, the source compartment produces a membrane-bound packet of
molecules called a vesicle.
After release, this vesicle specifically recognizes the correct target
compartment within the cell, and fuses with it~\cite{alberts2013essential}.
A lot of information about the molecules that form the machinery of
the VTS has been discovered, including their regulatory interaction
with each other~\cite{bonifacino2004mechanisms}. 
%
In spite of this detailed knowledge at the level of the molecules, the
structure of the VTS network, or the road-map of the eukaryotic cell,
is far from complete.
For example, although the localization of various SNAREs --a class of
molecules that participate in the control of VTS-- in the cell is
known, and also their site of action~\cite{hong2014tethering}, for
most SNAREs, how they first reached the compartments they reside in is
not known.
The current knowledge of the network is put together from a patchwork
of biological experiments and is scattered across several publications.
Even after this information is collected and put together, we find
that the network obtained is still not complete; new vesicles and new
contents in previously known vesicles are constantly being discovered
(some new discoveries include~\cite{chanaday2017you, d2017tethering,
rodepeter2017indication, zhao2017conserved}).
The synthesis for the unknown pieces may be assisted by computation on
the graph model of VTSs.
%
%
%
%
%
%
%
In this paper, we are looking at the computational questions 
arising from the VTSs.

VTSs are regulated by the same molecules that they transport.
For the purpose of this paper, the VTS molecules we focus on are the transmembrane SNARE proteins.
SNAREs drive the recognition of the target compartment by vesicles and
their subsequent fusion.
The SNAREs can be divided into v-SNAREs (which are present
on vesicles) and t-SNAREs (which are present on compartments). A
vesicle fuses with a compartment if its v-SNARE can form a complex
with the t-SNARE present on that compartment.
Not all v- and t- SNARE combinations can form complexes; this
constraint forms part of the basis for the specificity of vesicle
traffic~\cite{jahn2006snares}.
%

%

We use the model of VTSs that has been presented
in~\cite{shukla2017discovering}.
Please look at Appendix~\ref{sec:model} for a detailed discussion on
pros and cons of the model.
We model the system as a labeled graph, where compartments are nodes
and transport vesicles are edges.
The molecular compositions of the compartments and vesicles are the
node and edge labels respectively.
The molecules can be active or inactive on any a compartment or
vesicle.
The activity states of molecules are also included in the labels.
Due to the biology of SNAREs of the VTSs our interest, a vesicle is enabled
by a set of {\em four} molecules such that one part of the set occurs in the
vesicle and the other part occurs in the target of the vesicle compartment.
The partition always divides the set in the set of three and one molecules.
%
%
%
%
%
The enabling molecules must be active in the vesicle
and target compartment respectively.
The pairs are called {\em fusing} sets and analogously
the vesicle is considered to be {\em fused} with the
destination compartment.
Not all sets of molecules can participate in the fusion; in the
biological cells, fusogenic SNARE complexes are discovered through
experiments.
%
%
Generally, the fusing pairs are found to be distinct for distinct vesicle-compartment fusions.
To ensure that a molecule that has participated in a fusion does not
interfere with fusion at compartments and vesicles, in the model, we
require that the molecule is inactive on appropriate compartments.
The activity of molecules is regulated by the other molecules, i.e.,
the presence and absence of the other molecules in a compartment or
vesicle may make the molecule active or inactive.
We call this regulation as {\em activity functions}.
%
%
%
%
The regulation controls are defined by a fusion pairing relation
containing pairs of molecules and activity
Boolean functions.

In the model, we assume that the system is in steady state and the
concentrations of the molecules in compartments do not change over
time.
Since our system is in steady state, we expect that any molecule that
leaves a compartment must come back via some path on the graph.
We call this property of VTS as {\em stability}.

%
As we have discussed earlier, our understanding of VTSs is partial.
%
%
%
%
The synthesis of the unknown pieces may be {\em assisted} by computation on
the graph model of VTSs.
In this paper, we consider several versions of the synthesis problem
involving different parts of VTSs that can be synthesized, such as
modifying labels, adding/deleting edges, and learning activity function.
We also consider variations on the properties against which we do
synthesis, namely stability, and $k$-connectedness that states that the VTS remains connected after removing any $k-1$ edges.
We have assumed that the given partial VTS is always well-fused whereas properties like stability and k-connectedness may not hold in the partial VTS.
In order to synthesize the parts of a VTS such that it satisfies the
constraints, we encode the synthesis problem into one of satisfiability of
quantified Boolean formulas(QBFs). 

We have implemented the encoding in a flexible tool,
which can handle a wide range of synthesis queries.
We have applied our tool on several VTSs including two found-in-nature
VTSs.
%

Our experiments suggest that some of the synthesis problems are
solvable by modern solvers and the synthesis technology may be useful
for biological research.
%


The rest of the paper is organized as follows.
%
In section~\ref{sec:prelim}, we present the graph model of VTSs and encoding of several
constraints on VTSs.
In section~\ref{sec:encoding}, we present the synthesis problems and their
encoding into QBF satisfiability.
In section~\ref{sec:experiments}, we present our implementation and experimental results.
We discuss related work in~\ref{sec:related} and conclude in section~\ref{sec:conclusion}.



\section{Preliminaries}
\label{sec:prelim}
In this section, we will present the model of VTS from~\cite{smtVTS}.
We will also present the constraints and properties on the VTSs, and their
encoding as a QBF formula.   
We model a VTS as a labelled graph along with assisting pairing matrices and
activating functions.

\begin{df}
  A VTS $G$ is a tuple $(\nodes,\mols,\edges,\nlabel,\pairs,\edgef,\nodef)$, where
  \begin{itemize}
  \item $\nodes$ is a finite set of nodes representing compartments in the VTS,
  \item $\mols$ is the finite set of molecules flowing in the system, 
  \item $\edges \subseteq \nodes \times (\powerset{\mols}-\emptyset) \times \nodes$ is the
    set of edges with molecule sets as labels,
  \item $\nlabel : \nodes \maps \powerset{\mols}$ defines the molecules present in the nodes,
  \item $\pairs \subseteq 2^{\mols}$ is pairing relation,
  \item $\nodef : \mols \maps \powerset{\mols} \maps \booleans$ is activity maps for nodes, and
  \item $\edgef : \mols \maps \powerset{\mols} \maps \booleans $ is activity maps for edges.
  \end{itemize}
\end{df}
$\nodes$, $\mols$, $\edges$, and $\nlabel$ define a labelled graph.
Additionally, $\pairs$ defines which molecules can fuse with which molecules,
and
$\nodef$ and $\edgef$ are the activity functions for molecules on
nodes and edges respectively.
The model captures the steady state of a VTS.
The analysis of the model will inform us about the network/graph
properties of VTSs.

%
A molecule $k$ is {\em active} at node $n$ if $k \in \nlabel(n)$ and
$\nodef(k,\nlabel(n))$ is true.
A molecule $k$ is {\em active} at edges $(n,M',n')$ if $k \in M'$ and
$\edgef(k,M')$ is true.
We call $G$ {\em well-structured} if molecules $M$ is divided into
two partitions $Q$ and $R$ such that
for each $P \in \pairs, |P \intersect Q| = 3 \land |P \intersect R| = 1 $, and
for each $(n,M',n') \in \edges$, $n \neq n'$ and
$M' \subseteq \nlabel(n) \intersection \nlabel(n')$.
In other words,
molecules are of two types $Q$ and $R$,
%
pairing relations have sets of four molecules such that three
are of one type and one is of another type
(motivated by the biochemistry of the fusion),
there are no self loops, and 
each edge carry only those molecules that are present in its source
and destination nodes.
An edge $(n,M',\_) \in \edges$ {\em fuses} with a node $n'$
if there are non-empty set of molecules $M'' \subseteq M'$ and $M''' \subseteq \nlabel(n')$
such that $M''$ are active in the edge, $M'''$ are active in $n'$, and $M'' \union M''' \in \pairs$.
We call $G$ {\em well-fused} if each edge $(n,M',n') \in \edges$ fuses
with its destination node $n'$
and can not fuse with any other node.


%

A {\em path} in $G$ is a sequence $n_1,...,n_\ell$ of nodes 
such that $(n_i,\_,n_{i+1}) \in \edges$ for each $ 0 < i < \ell$.
For a molecule $m \in M$,
an {\em $m$-path} in $G$ is a sequence $n_1,...,n_\ell$ of nodes 
such that $(n_i,M',n_{i+1}) \in \edges$ and $m \in M'$ for
each $ 0 < i < \ell$.
A node $n'$ is {\em ($m$-)reachable} from node $n$ in $G$ if there is a ($m$-)path
$n,...,n'$ in $G$.
%
%
We call $G$ {\em stable} if for each $(n,M',n') \in \edges$ and $m \in M'$,
$n$ is $m$-reachable from $n'$.
We call $G$ {\em connected} if for each $n,n' \in \nodes$,
$n'$ is reachable from $n$ in $G$.
We call $G$ $k$-connected if for each $\edges' \subseteq \edges$ and $|\edges'| < k$,
VTS $(\nodes,\mols,\edges-E',\nlabel,\pairs,\edgef,\nodef)$ is connected.

%

\subsection{Encoding VTS}

The conditions on the VTSs for a given size can be encoded as a QBF formula
with uninterpreted functions.
To encode the constraints, we need variables for each aspect of
VTS.
Let us suppose that the size of the graph is $\nu$ and a number of
molecules are $\mu$.
To fully finitize the problem, we also limit the maximum number $\pi$
of edges present between two nodes.
Here, we list the Boolean variables and uninterpreted function symbols
that encode parts of VTSs.
\begin{enumerate}

\item Boolean variable $n_{i,m}$ indicates if $m \in \nlabel(i)$
\item Boolean variable $e_{i,j,q}$ indicates if $q$th edge exists between $i$ and $j$.
\item Boolean variable $e_{i,j,q,m}$ indicates if $q$th edge between $i$ and $j$ contains $m$.
\item Boolean variable $p_{\{m_1,m_2,m_3,m_4\}}$ indicates if $\{m_1,m_2,m_3,m_4\} \in \pairs$
\item uninterpreted Boolean functions $\nodef_m : \booleans^\mu \maps \booleans$
encoding $\nodef(m)$ map
\item uninterpreted Boolean functions $\edgef_m : \booleans^\mu \maps \booleans$
encoding $\edgef(m)$ map
\end{enumerate}
We also have auxiliary Boolean variables that will help us encode the well-fused property. 
\begin{enumerate}
\item $a_{i,m}$ indicates that molecule $m$ is active at node $i$, i.e., $\nodef(m,L(i))$
  holds
\item $b_{i,j,q,m}$ indicates that molecule $m$ is active at $q$th edge $(i,M',j)$ between $i$ and $j$, i.e., $\edgef(m,M')$ holds
\end{enumerate}

We will describe several constraints that encode VTSs in this section.
In the next section, we will extend the encoding for the synthesis problem.
To avoid cumbersome notation, we will not explicitly write the ranges
of the indexing in the constraints.
$i$ and $j$ will range over nodes, i.e., from $1$ to $\nu$.
$m$ will range over molecules, i.e., from $1$ to $\mu$.
$q$ will range over edges between two nodes, i.e., from $1$ to $\pi$.

The following constraints encode the basic consistancy of VTSs.
\begin{align*}
  \texttt{EdgeC} =\;&\hspace{-1ex}\bigwedge\limits_{i,j,q} (\bigvee_m e_{i,j,q,m} )\limplies e_{i,j,q}
  \land
  \bigwedge\limits_{i,q} \neg e_{i,i,q}
  \land
  \bigwedge\limits_{\mathclap{i,j,q,m}} e_{i,j,q,m} \limplies (n_{i,m} \land n_{j,m} )
  \\
  \texttt{ActivityC} =\;&
  \bigwedge\limits_{\mathclap{i,j,q,m}} b_{i,j,q,m} \limplies e_{i,j,q,m} \quad\land\quad
  \bigwedge\limits_{i,m} a_{i,m} \limplies n_{i,m}
  \\
  \texttt{PairingC} =\;&
  \exists qr. \Land\limits_{\mathrlap{m_1,m_2,m_3,m_4}\;\;}
                         (p_{\{m_1,m_2,m_3,m_4\}} \limplies qr_{m_1} + qr_{m_2} + qr_{m_3} + qr_{m_4} = 3) 
  \\
  \texttt{Fusion1} =\;&
  \bigwedge\limits_{i,j,q} e_{i,j,q} \limplies
  \bigvee_{{m_1,m_2,m_3,m_4}} (\Land_{l=1}^4 ( b_{i,j,q,m_l} \lor a_{j,m_l} ) \land 
                        \Lor\limits_{l=1}^4 b_{i,j,q,m_l} \land\\
  & \qquad \qquad \qquad \qquad \Lor\limits_{l=1}^4 a_{j,m_l}\land p_{\{m_1,m_2,m_3,m_4\}})
  \\
  \texttt{Fusion2} =\;&
\bigwedge\limits_{\mathclap{i,j,q,m_1,m_2,m_3, l \in \{1,..,3\}}} b_{i,j,q,m_1} \land ..\land b_{i,j,q,m_l} \limplies \\
  & \hspace{2cm} \neg 
  \bigvee_{\mathclap{j \neq j^{\prime}, m_{l+1}^{\prime},..,m_{4}^{\prime}}} ( a_{j^{\prime},m_{l+1}^{\prime}} \land .. \land a_{j^{\prime},m_4^{\prime}} \land p_{\{m_1,..,m_l,m^{\prime}_{l+1},..,m^{\prime}_4\}})
  \\
  \texttt{Consistancy} =\;& \texttt{EdgeC} \land
  \texttt{ActivityC} \land \texttt{PairingC} \land
  \texttt{Fusion1} \land \texttt{Fusion2} 
\end{align*}
\texttt{EdgeC} states that each edge has at least one molecule,
there are no self loops, and edge labels are consistent with node labels.   
\texttt{ActivityC} states that active molecule are present.
\texttt{PairingC} states that all molecules are divided into two types
using $qr_m$ bit, which encodes if $m$ belongs to one type or another,
and any fusing set of molecules must have three molecules involved
from one type and one molecule from the other.
\texttt{Fusion1}, and \texttt{Fusion2} states the well-fused condition.
\texttt{Consistancy} is the conjunction of all of the above.

\paragraph{Activity functions}
We also need to encode that the activity of the molecules are
controlled by activity functions.
The input VTS may include concrete activity functions for some molecules,
and for the others the functions may be unknown and to be synthesized. 
The concrete functions can be given to us in many different ways,
for example as a lookup table, or a concise Boolean formula.
In the following section, we will assume the appropriate encoding is
used for the concrete functions and represent them by \texttt{NodeFun}$_m$
and \texttt{EdgeFun}$_m$ for node and edge regulations respectively.
We will use $\nodef_m$ and $\edgef_m$ to represent functions that
are unknown in a VTS.
%
%
Later we will be synthesizing the unknown activity functions and 
replace $\nodef_m$ and $\edgef_m$ with parameterized constraints that
encode a space of candidate functions.

\subsection{VTS properties}

For the synthesis of incomplete systems,
we need properties against which we synthesize the missing parts.
Here we will discuss two such properties proposed in earlier
works~\cite{smtVTS}, namely stability and $k$-connectedness.

\paragraph{Stability property}
We use Boolean variable $r_{i,j,m,p}$ to indicate if there is an
$m$-path from $i$ to $j$ of length less than or equal to $p$.
We use $m$-reachability to encode the stability condition in VTSs.
The following constraint recursively encodes that node $j$ is
$m$-reachable from node $i$ in less than $p$ steps.
Subsequently, we encode stability condition using the reachability variables.
\begin{align*}
  \texttt{Paths}(r) &= \bigwedge\limits_{\mathclap{i,j,m,p}} r_{i,j,m,p} \limplies (\bigvee_{q} \, e_{i,j,q,m} \lor \bigvee_{i\neq i^{\prime}} ( \, \bigvee_{q} e_{i,i^{\prime},q,m}) \land r_{i^{\prime},j,m,p-1} )
  \\
  \texttt{Loop}(r) &= \bigwedge\limits_{i,j,m} (\bigvee_{q} e_{i,j,q,m}) \limplies r_{j,i,m,\nu}
  \\
  \texttt{Stability} &= \exists r. \; \texttt{Paths}(r) \land \texttt{Loop}(r)
\end{align*}

\paragraph{$k$-connected property}
$k$-connectedness expresses robustness against failure of few edges.
Let us use $d_{i,j,q}$ to indicate $q$th edge between $i$ and $j$ is failed
and $r'_{i,j}$ to indicate if there is a path from $i$ to $j$ in
the modified VTS.
%
%
In the following, $\texttt{Fail}(d,k)$ encodes that only
existing edges can be failed and exactly $k-1$ edges are failed.
$\texttt{FReach}(d,r')$ defines reachability in the modified VTS.
We use a new variable $r'_{i,j,p}$ to encode reachability from
$i$ to $j$ in at most $p$ steps.
$\texttt{Connected}(r')$ says that all nodes are reachable from any
other node.
\begin{align*}
  \texttt{Fail}(d,k) = & 
  \bigwedge\limits_{i,j,q} d_{i,j,q} \limplies e_{i,j,q}  \land 
  \sum_{i,j,q} d_{i,j,q} = k-1\\
  \texttt{FReach}(d,r') = &\hspace{-1ex}
   \bigwedge\limits_{i,j,p}  \hspace{-1ex}r^{\prime}_{i,j,p} \hspace{-1ex}\limplies\hspace{-1ex} [\bigvee_{q} (e_{i,j,q} \land  \neg d_{i,j,q}) \hspace{-1pt}\lor \hspace{-2pt} (\bigvee_{\mathclap{i' \neq i}}  r^{\prime}_{i',j,p-1} \land  \bigvee_{q} (e_{i,i',q} \land \neg d_{i,i',q}) ]\\
  \texttt{Connected}(r') = & \Land\limits_{i,j} (r^{\prime}_{i,j,\nu} \lor r^{\prime}_{j,i,\nu})
\end{align*}
We will be synthesizing $k$-connected graphs.
We define $\texttt{Connected}(k)$ that says for all possible valid failures
the graph remains reachable. 
\begin{align*}
  \texttt{Connected}(k) = & \forall d.\;
          (\texttt{Fail}(d,k) \limplies \exists r'.\;\texttt{FReach}(d,r')
                                \land \texttt{Connected}(r'))
\end{align*}
Since $d$ variables in $\texttt{Connected}(k)$ are universally
quantified, $\texttt{Connected}(k)$ introduces quantifier alternations.
Therefore, synthesis against this property will require QBF reasoning.
We may make the formula quantifier free  by considering all possible failures
separately and introducing a vector of reachability variables for each
failure.
However, this will blow up the size of the formula and may not be
solvable by a SAT solver.
%



\section{Synthesis for VTS}
\label{sec:encoding}
In this section, we will present a list of synthesis problems that may
arise from the partially available information about a VTS and our synthesis method
for the problems.

\subsection{Problem Statements}

We will assume that we are given a VTS, whose all components
are not specified.
Our objective is to find the missing parts.
The missing parts can be in any of the components of VTS. 
For example, some undiscovered edges or nodes, or insufficient
knowledge about the presence of molecules in some part of the VTS.
To cover most of the likely variations of this missing information,
we have encoded the following variants of VTS synthesis problem.

\begin{enumerate}
\item Fixing VTS by adding edges 
\item Fixing VTS by adding molecules to the labels
\item Fixing VTS by learning activity functions
\item  Fixing VTS by both adding/deleting parts
\end{enumerate}

\subsection{Encoding Incomplete VTS}

In our synthesis method, we take a VTS $G =
(\nodes,\mols,\edges,\nlabel,\pairs,\edgef,\nodef)$ as input.
We allow activity functions not to be specified.
We construct the following constraints to encode the available information
about $G$.
We encode both the present and the absent components in $G$.
Later, the constraints will help us encode the synthesis problems.
\begin{align*}
  \texttt{PresentE} &= \land \{e_{i,j,q,m}|(i,M_1,j),...,(i,M_{q'},j) \in \edges \land q \leq q' \land m \in M_q \}\\
  \texttt{PresentN} &= \land \{n_{i,m}| m \in \nlabel(i) \land i \in \nodes \}\\
  \texttt{PresentP} &= \land \{p_{\{m_1,m_2,m_3,m_4\}}| \{m_1,m_2,m_3,m_4\} \in \pairs \} \\
  \texttt{KnownActiveN} &= \land \{ a_{i,m} = \texttt{NodeFun}_m (n_{i,1},\dots,n_{i,\mu}) | \nodef_m \text{ is defined.} \} \\
  \texttt{KnownActiveE} &= \land \{ b_{i,j,q,m} = \texttt{EdgeFun}_m(e_{i,j,q,1}, .., e_{i,j,q,\mu})
                   | \edgef_m \text{ is defined.} \} \\
  \texttt{PresentCons} & = \texttt{PresentE} \land \texttt{PresentN} \land 
  \texttt{PresentP} \land \texttt{KnownActiveE} \land \\
  &  \texttt{KnownActiveN}
\end{align*}
We also collect the variables that are not set to true in $\texttt{PresentCons}$.
\begin{align*}
  \texttt{AbsentELabel} &=
  \{e_{i,j,q,m}|(i,M_1,j),...,(i,M_{q'},j) \in \edges \land 
                          0 < q \leq q' \land m \not\in M_{q} \}\\
  \texttt{AbsentE} &= \{e_{i,j,q}|(i,M_1,j),...,(i,M_{q'},j) \in \edges \land 
                    q' < q \leq \pi \}\\
  \texttt{AbsentNLabel} &= \{n_{i,m}| m \not\in \nlabel(i) \land i \in \nodes \}\\
  \texttt{AbsentP} &= \{p_{\{m_1,m_2,m_3,m_4\}}| {\{m_1,m_2,m_3,m_4\}} \not\in \pairs \} \\
  \texttt{UnknownActive} &=  \Land \{ a_{i,m} = \nodef_m (n_{i,1},\dots,n_{i,\mu}) | \nodef_m \text{ is undefined.} \} \union \\
   &\quad\quad\quad  \{ b_{i,j,q,m} = \edgef_m(e_{i,j,q,1}, .., e_{i,j,q,\mu})
                   | \edgef_m \text{ is undefined.} \}
\end{align*}
We have defined $\texttt{AbsentELabel}$, $\texttt{AbsentE}$, $\texttt{AbsentN}$, and
$\texttt{AbsentP}$
as sets.
They will be converted into formulas depending
on the different usage in the synthesis problems. 


\subsection{Encoding synthesis property}

We will do synthesis against the following property that says the VTS
is stable and 3-connected.
%
%
\begin{align*}
  \texttt{Property} =  \texttt{Stability} \land \texttt{Connected}(3) 
\end{align*}
The property was proposed in~\cite{shukla2017discovering}.
However, the biological relevance of the property is debatable and open for change.
Our tool is easily modifiable to support any other property that may be deemed 
interesting by the biologists.

\subsection{Encoding synthesis constraints}

Now we will consider the encodings for the
listed synthesis problems.
The presented variations represent the encodings
supported by our tool.
Additionally, the combinations of the variation are also possible and
our tool easily supports them.
For simplicity of the presentation, we assume that if we are
synthesizing an aspect of VTS, then all other aspects are fully given.
Therefore, we will describe two kinds of constraints for synthesis
problems.
One will encode the variable part in the synthesis problem and
the other encodes the fixed parts.
Subsequently, the two constraints will be put together with 
$\texttt{Consistancy}$ and 
$\texttt{Property}$ to construct the constraints for synthesis.

\subsubsection{Fixing VTS by adding edges}
Now we will consider the case when we add new edges to VTS to satisfy the properties.
In the following, the pseudo-Boolean formula $\texttt{AddE}$ encodes
that at most $slimit$ new undeclared edges may be added in the VTS.
\texttt{FixedForEdge} encodes the parts of the VTS that are not allowed to change.
\begin{align*}
  &\texttt{AddE}(slimit) = \sum~\texttt{AbsentE} \leq slimit\\
  &\texttt{FixedForEdge} = \texttt{PresentCons} \land \texttt{UnknownActive} \land\\
   & \quad\quad\quad\quad \lnot \Lor \texttt{AbsentELabel} \;\union\;
                    \texttt{AbsentNLabel} \;\union\;
                    \texttt{AbsentP}
\end{align*}
We put together the constraints and obtain the following formula.
\begin{align*}
  & \texttt{SynthE}(slimit) =
       \texttt{Consistancy}\land \texttt{Property} \land
   \texttt{FixedForEdge} \land
  \texttt{AddE}(slimit)
\end{align*}
Similar to what we have seen $\texttt{Consistancy}$ encodes the basic constraints about VTS,
$\texttt{Property}$\; encodes the goal, and
the rest two are defined just above.
A satisfying model of $\texttt{SynthE}$ will make 
some of the edges in $\texttt{AbsentE}$ true such that~$\texttt{Property}$ is satisfied.
We limit the addition of the edges, since we look for a fix that require minimum number
of changes in the given VTS.
We start with $slimit = 1$ and grow one by one until $\texttt{SynthE}(slimit)$
becomes satisfiable.

In the later synthesis problems, we will construct a similar QBF
formula with same first two parts and the last two are due the
requirements of the synthesis problem.

\paragraph{\bf Fixing VTS by adding molecules to the labels:}
The system may also be fixed only by modifying labels on the edges or the nodes instead
of adding edges.
Here let us consider only adding molecules to the labels of edges.
In the following, the formula encodes that only $slimit$ edge labels may be added.
\begin{align*}
  &\texttt{AddLabelEdge}(slimit) = 
    \sum~\texttt{UnknwonEdgeLabel}  \leq slimit\\
  & \texttt{FixedForLabel} = \texttt{PresentCons} \land \texttt{UnknownActive} \land\\
  & \quad \quad \lnot \Lor \texttt{AbsentE} \;\union\;
                    \texttt{AbsentNLabel} \;\union\;
                    \texttt{AbsentP}\\
  &\texttt{SynthLabel}(slimit) = \\
  & \quad\quad
    (  \texttt{Consistancy}\land \texttt{Property} \land
   \texttt{FixedForLabel} \land \texttt{AddLabelEdge}(slimit) )
\end{align*} 
Similar to the previous encoding, we solve the
satisfiability of the above formula to obtain additional molecules
that may be added to the edge labels to satisfy the properties.

\subsubsection{Fixing VTS by learning activity functions:}

Now we consider a scenario where some of the activity functions for
some of the molecules are missing.
The activity functions are $\mu$-input Boolean functions.
First, we choose a class of formulas for the candidate functions.
We encode the candidates in a formula with parameters.
By assigning different values for the parameters, a solver may select
different candidates for the activity functions.
We will illustrate only one class of formulas.
However, we support other classes of formulas, for example, $k$-CNF.

In the following, the formula \texttt{NNFTemplate} encodes a set of
negation normal form functions that take $y_1,..,y_\mu$ as input and
contain $\lambda$ literals.
We use $\texttt{Gate}$ to encode a gate that takes a parameter integer
$x$ to encode various gates.
We use $\texttt{Leaf} $ to encode the literal at some position.
Both are stitched to define \texttt{NNFTemplate}.
To encode the set of NNF formulas with $\lambda$ literals, it has
finite-range integer variables
$z_1,..,z_{2\lambda}$ as parameters.


\begin{align*}
  &\texttt{Gate}( x, w_1, w_2 ) = ( x  = 1 \limplies w_1  \land w_2 ) \land  
  ( x = 2  \limplies w_1  \lor w_2 ) \\
  &\texttt{Leaf}( x, [y_1,..,y_\mu] ) =
  \Land\limits_{l=1}^{\mu} ( x = 2l-1  \limplies y_{l}) \land ( x = 2l  \limplies \lnot y_{l})\\
  &\texttt{NNFTemplate}([z_1,..,z_{2\lambda}],[y_1,..,y_\mu] ) = \\
  &\exists w_1,..,w_{2\lambda}.\;w_1 \land \Land_{l=1}^{\lambda} w_{\lambda+l} = \texttt{Leaf}( z_{\lambda+l}, [y_1,..,y_\mu] ) \land
  w_{l} = \texttt{Gate}( z_l, w_{2l},w_{2l+1})
\end{align*}

Using the template we define the constraints $\texttt{FindFunctions}(z,\lambda)$
that encodes the candidate functions that satisfy the activity requirements,
where $z$ is the vectors of parameters for encoding parameters
for each molecule,
and $\lambda$ limits the size of the candidate functions. 
We fix the all other aspects of the VTS to be fixed via constraints
$\texttt{FixedForFunctions}$.
\begin{align*}
  &\texttt{FindFunctions}(z,\lambda) =\\
  &\Land \{ \bigwedge\limits_{i} n_{i,m} \limplies a_{i,m} = 
  \texttt{NNFTemplate}([z_{m,1},..,z_{m,2\lambda}],[n_{i,1},\dots,n_{i,\mu}] ) \\
  & \hspace{8cm}|
  \nodef_m \text{ is undefined}\} \\
  &\Land \{ \bigwedge\limits_{i,j,q} e_{i,j,q,m} \limplies b_{i,j,q,m} = 
  \texttt{NNFTemplate}([z_{i,j,q,1},..,z_{i,j,q,2\lambda}],[e_{i,j,q,1},..,e_{i,j,q,\mu}] ) \\
  & \hspace{8cm} | \edgef_m \text{ is undefined}
  \}\\
  & \texttt{FixedForFunctions} = \texttt{PresentCons} \land\\
  & \quad \quad
\lnot \Lor \texttt{AbsentE} \;\union\; \texttt{AbsentELabel} \;\union\;
                    \texttt{AbsentNLabel} \;\union\;
                    \texttt{AbsentP}\\
  &\texttt{SynthFunction}(z,\lambda) = \\
  & \quad
    (  \texttt{Consistancy}\land \texttt{Property} \land
   \texttt{FixedForFunctions} \land \texttt{FindFunctions}(slimit) )
\end{align*}
We construct $\texttt{SynthFunction}(z,\lambda)$ similar to the earlier
variations.
By reading of the values of $z$ in a satisfying model of the formula,
we learn the synthesized function.

\subsection{Fixing VTS by both adding/deleting parts:}
Now we will consider repairing of VTS by allowing not only addition but also
deletion of the molecules, edges, functions, or pairing matrix.
We have encoded the repairing in our tool by introducing flip bits
for each variable that is modifiable in the VTS.
We illustrate the repairing on one class of variables and rest can be
easily extended.
Let us consider repairing of node labels.
For each bit $n_{i,m}$, we create a bit $flip_{i,m}$.
We add constraints that take xor of VTS assigned values for  $n_{i,m}$
and $flip_{i,m}$.
We also limit the number of $flip_{i,m}$ that can be true, therefore
limiting the number of flips.
The above constraints are encoded in $\texttt{FlipN}(slimit)$.
\begin{align*}
    &\texttt{FlipN}(slimit,flip) = \Land \{n_{i,m}\lxor flip_{i,m}| m \in \nlabel(i) \land i \in \nodes \} \land \\
  & \quad \quad \quad
    \Land \{ \lnot n_{i,m}\lxor flip_{i,m}| m \not \in \nlabel(i) \land i \in \nodes \} \land  \sum\limits_{i,m} flip_{i,m} \leq slimit
\end{align*}
Similar to the earlier variations, we construct
$\texttt{SynthRepairNode}(slimit)$ for the repair.
In that, $\texttt{FixedForNodeRepair}$ encodes all the parts of VTS that do not change.
\begin{align*}
    & \texttt{FixedForNodeRepair} =  \texttt{PresentE} \land  
      \texttt{PresentP} \land \texttt{KnownActiveE} 
      \land \\
  & \quad \texttt{UnknownActive} \land \texttt{KnownActiveN} \land \lnot 
    \Lor \texttt{AbsentE} \;\union\; \texttt{AbsentNLabel} \;\union\;
    \texttt{AbsentP}\\
  &\texttt{SynthRepairNode}(slimit,flip) = \\
  & \quad\quad
    (  \texttt{Consistancy}\land \texttt{Property} \land
    \texttt{FixedForNodeRepair} \land \texttt{FlipN}(slimit,flip) )
\end{align*}
A satisfying model of $\texttt{SynthRepairNode}(slimit,flip)$ will assign some
$flip$ bits to true.
We will learn from the assignments the needed modifications in the VTS. 


\section{Implementation and Experiments}
\label{sec:experiments}
\begin{sidewaysfigure}[t]
  \centering
  \begin{tabular}[t]{|c@{}|@{}c@{}|@{}c@{}|@{}c@{}|@{}c@{}|@{}c@{}|@{}c@{}|@{}c@{}|@{}c@{}|@{}c@{}|@{}C{4cm}@{}|}\hline
    {\multirow{2}{*} \textbf{}}  & \multicolumn{2}{c|}{\textbf{Add}} & \multicolumn{2}{c|}{\textbf{Add}} & \multicolumn{2}{c|}{\textbf{Learning NNF}}  &  \multicolumn{2}{c|}{\textbf{Learning}} &  \multicolumn{2}{c|}{\textbf{Add/Delete}} \\
    {\multirow{2}{*} \textbf{Table a}}  & \multicolumn{2}{c|}{\textbf{edge}} & \multicolumn{2}{c|}{\textbf{molecules}} & \multicolumn{2}{c|}{\textbf{(only $\land$ and $\lor$)}}  &  \multicolumn{2}{c|}{\textbf{k-CNF}} &  \multicolumn{2}{c|}{\textbf{parts}} \\
    \cline{2-11}
    {} & {\textbf{Time}} & {\textbf{\#C}} & {\textbf{Time}} & {\textbf{\#C}} & {\textbf{Time}} & {\textbf{\#C}} & {\textbf{Time}} & {\textbf{\#C}} & {\textbf{Time}} & {\textbf{\#C}} \\
    \hline
    
    plos1-dia[3C]& 0.326 &$\infty$& 0.312 &$\infty$& 0.669 & $\infty$ & 0.966 &$\infty$& 0.277 & -1 E, -1 AE, -1 AN. +1 E, +1 N. \\\hline
    plos2-dia[4C] & 0.266 & 0   & 0.322 & 0  & 1.409  & 0 & 2.114 & 0 &  0.337 & 0 \\\hline
    sub-mammal[3C]  & 0.767 & 1 E  & 1.049 & 5 PE & 3.523 & 1E & 4.961 & 1E & 1.172  & -1 E, -2 PE, -1 AN. +1 E, +4 PE, +4 N, +2 AN, +2 AE. \\\hline
    node4[3C]  & 1.554  & 1 E   &  3.859 & 12 PE  &  5.286  & $\infty$ & 4.502 &$\infty$& 2.194  & -2 E, -2 PE, -1 N, -1 AN, -1 AE. +12 N, +8 E, +1 PE.\\\hline
    yeast-graph[3C]   & 95.016    & 2 E  &   timeout  & N/A   & 1571.42  & 2 E  & 530.210   & 2 E & 72.316  &  -1 E, -1 N, -1 AE, -1 AN, -1PE. +2 E, 7 PE, 8 N. \\\hline

mammal-graph[3C]  &  timeout     & N/A  &  timeout     & N/A    &  timeout         & N/A      &  timeout    &  N/A    &  timeout     & N/A\\\hline
  \end{tabular}
  \begin{tabular}[t]{|c@{}|@{}c@{}|@{}c@{}|@{}c@{}|@{}c@{}|@{}c@{}|@{}c@{}|@{}c@{}|@{}c@{}|@{}c@{}|@{}C{4cm}@{}|}\hline
    {\multirow{2}{*} \textbf{}}  & \multicolumn{2}{c|}{\textbf{Add}} & \multicolumn{2}{c|}{\textbf{Add}} & \multicolumn{2}{c|}{\textbf{Learning NNF}}  &  \multicolumn{2}{c|}{\textbf{Learning}} &  \multicolumn{2}{c|}{\textbf{Add/Delete}} \\
    {\multirow{2}{*} \textbf{Table b}}  & \multicolumn{2}{c|}{\textbf{edge}} & \multicolumn{2}{c|}{\textbf{molecules}} & \multicolumn{2}{c|}{\textbf{(only $\land$ and $\lor$)}}  &  \multicolumn{2}{c|}{\textbf{k-CNF}} &  \multicolumn{2}{c|}{\textbf{parts}} \\
    \cline{2-11}
    {} & {\textbf{Time}} & {\textbf{\#C}} & {\textbf{Time}} & {\textbf{\#C}} & {\textbf{Time}} & {\textbf{\#C}} & {\textbf{Time}} & {\textbf{\#C}} & {\textbf{Time}} & {\textbf{\#C}} \\
    \hline
    
    plos1-dia & 0.041&$\infty$& 0.320 &$\infty$& 0.225 & $\infty$ & 0.33&$\infty$& 3.74 & -1 E, -1 PE, - 1 N, -1 PE. +1 AE, +1 PE, +1 N\\\hline
    plos2-dia & 3.97 & 0 &  2.647 & 0  & 5.941 & 0 & 5.680 & 0 & 3.56 & 0 \\\hline
    sub-mammal & 3.483 & 1 E  & 4.379 & 5 PE  & 29.980 & 1 E  & 10.405 & 1 E & 3.650  & -1 E, -2 PE, -1 AN. +1 E, +4 PE, +4 N, +2 AN, +2 AE \\\hline
    node4  & 4.150  & 1 E  & 10.562  & 12 PE & 3.401  & $\infty$ & 4.760 &$\infty$&  5.05  & -2 E, -2 PE, -1 N, -1 AN, -1 AE. +12 N, +8 E, +1 PE \\\hline
    yeast-graph & 40.225  & 2 E  &   timeout  & N/A   & 1393.84  & 2 E  & 468.161   & 2 E & 69.81  &  -1 E, -1 N, -1 AE, -1 AN, -1PE. +2 E, 7 PE, 8 N. \\\hline
    mammal-graph   &  timeout     & N/A  &  timeout     & N/A    &  timeout         & N/A      &  timeout    &  N/A   &  timeout     & N/A\\\hline
  \end{tabular}
    \caption{Run-times for synthesis queries. \#C stands for minimum changes in the synthesized VTS in comparison with the given partial VTS. Time is reported in seconds. (a) The solver used is DepQBF (b) The solver used is Z3. The sub-mammal is a subgraph of the complete mammal-graph. In the Add/Delete parts column, ‘+’n sign is used to show the addition of n number of the molecules, similarly ‘-’n is used to show the removal of n number of molecules. In the table, N is node labels, AN is active node molecules, E is edges, PE is molecule presence on the edge and AE is active molecules on the edge. The [kC] stands for k graph connectedness which is part of only DepQBF experiments.}

  \label{tab:qf-graph}
\end{sidewaysfigure}


We have implemented the encodings in a tool
called~\ourtool\footnote{{\url{https://github.com/arey0pushpa/pyZ3}}}.
The tool takes a partially defined VTS as input in a custom designed
input language.
The input is then converted to the constraints over VTS. 
The tool can not only synthesize the above-discussed queries, but also their
combinations.
For example, our tool can modify labels of nodes or edges while
learning activity functions.
Our tool is developed in C++ and uses~\zthree~\cite{z3} infrastructure for
processing formulas. 
Since some of the formulas involve alternation of quantifiers over
Boolean variables Z3 is not a suitable choice for those examples.
We translate the formulas created by Z3 tool into a standard
QDIMACS~\cite{qdimacs} format and use as an input for QBF solvers. 
We use~\depqbf~\cite{lonsing2010depqbf} for solving of QBF formulas. 
Our tool includes about 7000 lines of code.

We have applied~\ourtool~on six partially defined VTSs.
The results are presented in table~\ref{tab:qf-graph} for both the solvers
\depqbf and \zthree.
To use~\zthree, we remove \texttt{Connected} constraints, such that the queries becomes
quantifier-free.
%
%
The experiments were done on a machine with Intel(R) Core(TM) i3-4030U
CPU @ 1.90GHz processor and 4GB RAM with 30 min (1800 sec) timeout.
%
The first four VTSs are synthetic but inspire from literature for
typical motifs in VTSs. 
The third VTS is a subgraph of the last VTS.
The fifth VTS is taken from~\cite{burri2004complete}.
The last VTS represent mammalian SNARE map created by studying the literature references.  

The table shows timing for various synthesis queries.
For each synthesis query, we have two columns.
One column reports the timing and the other reports the minimum changes
needed to obtain a valid VTS.
$\infty$ indicates that any number of changes with the synthesis query
search space can obtain the VTS.
%
In the table, we are reporting five synthesis queries
The first one only adds new labelled edges to the graph.
We have ranked the all possible graph edits with the simple rank of
minimum updates.
%
The second query adds new labels to the edge.
%
The third query synthesizes NNF Boolean functions only containing
$\land$ and $\lor$ gates for activity functions, while allowing
more edges to be added.
The result shows the basic template of 4 leaves and 3 gates.
%
%
To illustrate the versatility of our tool, the fourth query
synthesizes $3$-CNF functions (encoding not presented).
%
Finally, we report queries that allows both addition and deletion of edges, and labels
of node and labels.

%
%


\section{Related Work}
\label{sec:related}
In recent years, there has been a wide range of methods
developed for the similar synthesis problems~\cite{sketch,sygus,exampleSynth}.
They range from filling gaps an implementation of C programs from the pool of template predicates to learn a program from example
runs of the program.
In the course of developing such methods,
the background technology, i.e. solving of quantified
constraints has been evolving rapidly~\cite{lonsing2010depqbf,z3Quant}.

There has been some work in
applying synthesis technique in biology especially in gene regulatory networks~\cite{shavit2016automated, fisher2015synthesising}. A very recent work~\cite{fisher2015synthesising} synthesize executable gene regulatory networks from single-cell gene expression data. 
Synthesis technique is also used in optimal synthesis for chemical reaction networks~\cite{cardelli2017syntax}. The~\cite{fisher2015synthesising} uses constraint (satisfiability) solving techniques for the synthesis whereas ~\cite{shavit2016automated} uses SMT for synthesis. The paper~\cite{cardelli2017syntax} in addition to using SMT over ODE, uses a template-guided approach. In our case queries contain quantifiers so we have employed QBF solving with Z3 for the solving the synthesis problem. To our best knowledge, this is the first application of synthesis in VTS.

\section{Conclusion}
\label{sec:conclusion}
In this paper, we presented encodings of the synthesis problems
that may arise from VTSs.
We demonstrated that our tool based on the encodings
scale up to the relevant sizes of the VTSs for some synthesis queries.
Our tool timed out on larger examples.
We are working to improve the performance of our tool.
We will take this tool to the biologists and develop wet experiments that may validate some synthesis results from the tool.
Our model of VTSs is static graphs.
In future, we will study the dynamic behaviors of VTSs.
It will allow us to predict behaviors after the perturbations in the VTSs
and more ways to test the predicted synthesis results.


\bibliographystyle{unsrt}
\bibliography{main}

\appendix
\clearpage

\section{Discussion on the choice of VTS model}
\label{sec:model}
The molecules transported by the VTS are themselves its regulators. The molecules in a
compartment/vesicle may be active or inactive. The molecules that are responsible for vesicle fusion
are called SNARE proteins~\cite{jahn2006snares, wickner2008membrane}. Active SNAREs present on vesicles (v-SNAREs) bind with their
cognate active SNAREs on the target compartment (t-SNAREs) to enable vesicle fusion. A cell
contains multiple kinds of v- and t-SNAREs. Only specific pairs of v and t SNAREs can bind to
each other and participate in fusion. Fusion compatible v- and t- SNAREs are determined by
biological experiments. Different vesicle-compartment fusions in the cell are brought about by
different v- and t-SNARE pairs. A molecule that participates in a given fusion reaction must not
interfere with fusion at different compartments or vesicles. Therefore, SNAREs must be kept in an
inactive form in appropriate compartments/vesicles. The activity of molecules is regulated by the
other molecules, i.e., the presence and absence of the other molecules in a compartment or vesicle
may make the molecule active or inactive. We call this regulation as activity functions.
In the VTS model, we assume that the system is in steady state and the concentrations of the
molecules in the compartments do not change over time. We define SNARE pairing specificity by a
fusion pairing relation containing pairs of SNAREs and molecular regulation by activity Boolean
functions. Since the system is in steady state, we expect that any molecule that leaves a
compartment must come back via some path on the graph. We call this property of VTS as stability.

Our model is inspired by~\cite{shukla2017discovering}. On the timescales of minutes, our following assumptions reasonably capture the important aspects of the Rothman-Schekman-Sudhof (RSS) model~\cite{rothman2002machinery} of vesicle traffic system.
\begin{enumerate}
\item A cell is a set of compartments exchanging vesicles.
\item Compartments are neither created nor destroyed.
\item Each compartment is in steady state, gain and loss balance.
\item Molecules are neither created nor destroyed.
\item Molecules move via vesicles of uniform size.
\item Identical vesicles have identical target compartments.
\item Fusion of vesicles to compartments is driven by specific SNARE pairing.
\item The activity of a SNARE can be regulated by other molecules present on the same compartment
or vesicle.
\item An active SNARE pair is necessary and sufficient for fusion. 
\end{enumerate}
 
SNARE proteins are the agents of vesicle fusion in eukaryotic cells. When SNAREs on vesicles (v-SNAREs) encounter their cognate SNAREs on target compartments (t-SNAREs), they form SNARE complexes~\cite{jahn2006snares}, and a single SNARE complex releases enough energy to enable membrane fusion~\cite{van2010one}. SNAREs are identified by the presence of a conserved 60-70 stretch of amino acids called the SNARE motif. Based on amino acid sequence, SNARE motifs fall into 4 classes: Qa, Qb, Qc, and R~\cite{jahn2006snares}. Across all intracellular vesicle fusion reactions, the associated SNARE complexes contain one of each of the four kinds of SNARE motifs; the v-SNARE contributes a single SNARE motif, usually it is an R-SNARE (although, exceptions are known: Sec22b and Ykt6 are both R-SNAREs which form parts of t-SNAREs~\cite{hong2005snares}) and the rest of the three SNARE
motifs are contributed by the t-SNARE. In the cell, different vesicle fusion reactions are associated with distinct v- and t-SNARE pairs.

The paper~\cite{shukla2017discovering} consider three Q SNARES as a single molecule,  we have extended this model by considering each complex molecule as distinct. In contrast to the~\cite{shukla2017discovering}, we allow Q and R-SNARE type distribution across the whole system to be uneven. In our model fusion is driven by an active combination of three Q SNARE and one R SNARE molecule. We have relaxed the pairing matrix constraint to comply with this fact. For biological efficiency and optimality reasons, we do not allow self-edges to be present in the VTS. 



\section{The Natural VTSs}
\label{sec:ex-vts}
Here we will present the two VTS collected from the literature.

\subsection{Mammalian VTS}
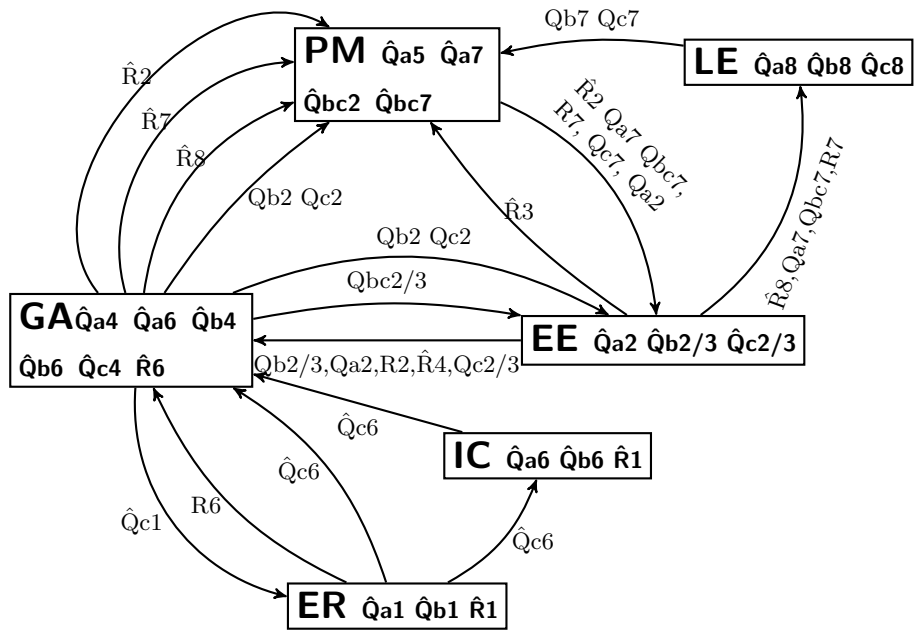
\begin{figure}[t]
  \centering
  \begin{tikzpicture}[->,>=stealth',auto,node distance=5cm,
  thick,main node/.style={rectangle,draw,font=\sffamily\Large\bfseries}]
  \node[main node,text width=3cm] (ga) {GA{\small \^{Qa4} \^{Qa6} \^{Qb4} \^{Qb6} \^{Qc4} \^{R6}}};
  \node[main node] (ic) [below right of=ga,yshift=20mm,xshift=20mm] {IC {\small \^{Qa6} \^{Qb6} \^{R1}}};
  \node[main node] (er) [below right of=ga] {ER {\small \^{Qa1} \^{Qb1} \^{R1}}};
  \node[main node,text width=2.5cm] (pm) [above right of=ga] {PM {\small \^{Qa5} \^{Qa7} \^{Qbc2} \^{Qbc7}}};
  \node[main node] (ee) [below right of=pm] {EE {\small \^{Qa2} \^{Qb2/3} \^{Qc2/3}}};
  \node[main node] (le) [above of=ee,yshift=-13mm,xshift=18mm] {LE {\small \^{Qa8} \^{Qb8} \^{Qc8}}};

  \path (ic) edge node [below] {\^{Qc6}} (ga);
  \path (er) edge[bend right=20] node [right] {\^{Qc6}} (ic);

  \path (er) edge[bend right=20] node [left] {\^{Qc6}} (ga);
  \path (er) edge[bend left=20] node [left] {R6} (ga);
  \path (ga) edge[bend right=40] node [left] {\^{Qc1}} (er);

  \path (ga) edge[bend left=25] node [above] {Qb2 Qc2} (ee);
  \path (ga) edge[bend left=10] node [above] {Qbc2/3} (ee);
  \path (ee) edge[bend left=0] node [below] {Qb2/3,Qa2,R2,\^{R4},Qc2/3} (ga);

  \path (le) edge[bend right=10] node [above] {Qb7 Qc7} (pm);
  \path (ee) edge[bend right] node [below,rotate=70] {\^{R8},Qa7,Qbc7,R7} (le);

  \path (ee) edge[bend left=10] node [above] {\^{R3}} (pm);
  \path (pm) edge[bend left=30] node [above,rotate=-45,text width = 2.5cm] {\^{R2} {Qa7} Qbc7, R7, Qc7, Qa2} (ee);

  \path (ga) edge[bend left=10] node [right] {{Qb2} Qc2} (pm);
  \path (ga) edge[bend left=80] node [above] {\^{R2}} (pm);
  \path (ga) edge[bend left=52] node [above] {\^{R7}} (pm);
  \path (ga) edge[bend left] node [above] {\^{R8}} (pm);
  \end{tikzpicture}
  \caption{A found-in-nature VTS. Nodes and edges are labelled with sets of molecules. \^{} indicates that the molecule is active.}
  \label{fig:mukund-vts}
\end{figure}


The figure~\ref{fig:mukund-vts} represent mammalian SNARE map
created by studying the wide array of literature.
To construct the map, we have assumed that vesicles only contain a
single active v-SNARE, and we have attributed t-SNAREs and inactive
v-SNAREs that travel between compartments to one of the known vesicles
that go between the same source and target compartments.
In order to identify the active SNARE complex involved in any
particular vesicle fusion, we used two criteria.
The SNARE complex is formed \textit{in vivo}. In most papers, this is
determined by immunoprecipitation of the SNARE complex from the
relevant cell fraction.
Blocking SNARE complex formation (for example, using antibodies
against these SNAREs, or using cytosolic forms of these SNAREs) blocks
the specific transport step.
Note that these vesicles have been collected from multiple cell types, and
any given cell type is likely to contain only a subset of the vesicles in
the map.

In this figure, the rectangles represent compartments, the identities
of compartments are written within ER=endoplasmic reticulum,
ERGIC=ER-Golgi intermediate compartment, RE=recycling endosome,
EE=early endosome, LE=late endosome, LYS=lysosome, PM=plasma
membrane. The arrows represent vesicle edges.


\subsection{Yeast VTS}

\begin{figure}[t]
  \centering
  \begin{tikzpicture}[->,>=stealth',auto,node distance=4.5cm,
  thick,main node/.style={rectangle,draw,font=\sffamily\Large\bfseries}]
  \node[main node,,text width=2.5cm] (golgi) [] {GOLGI {\small \^{Qa2} \^{Qb2} \^{Qc2} \^{Qa5} \^{Qb5} \^{Qc5}}};
  \node[main node,text width=2.5cm] (pvac) [right of=golgi] {PRE-VAC {\small \^{Qa3} \^{Qb2} \^{R3} \^{R4}}};
  \node[main node,text width=2cm] (vac) [right of=pvac] {VAC {\small \^{Qa4} \^{Qb2} \^{R3} \^{R4}}};
  \node[main node,text width=2.5cm] (pm_up) [above right of=golgi] {PM {\small \^{Qa1} \^{Qbc1}}};
  \node[main node] (pm_down) [below right of=golgi] {PM {\small \^{Qa6} \^{Qb6} \^{R6} \^{R4}}};

  \path (golgi) edge[bend right] node [right] {\^{R1}} (pm_up);
  \path (pm_up) edge[bend right] node [right] {\^{R1}} (golgi);

  \path (golgi) edge[bend right] node [right] {\^{Qc6}} (pm_down);
  \path (pm_down) edge[bend right] node [right] {\^{R1}} (golgi);

  \path (golgi) edge[] node [below] {\^{Qc3}} (pvac);
  \path (pvac) edge[bend left=20] node [above] {\^{Qc4}} (vac);

  \end{tikzpicture}
  \caption{Yeast VTS}
  \label{fig:yeast-vts}
\end{figure}
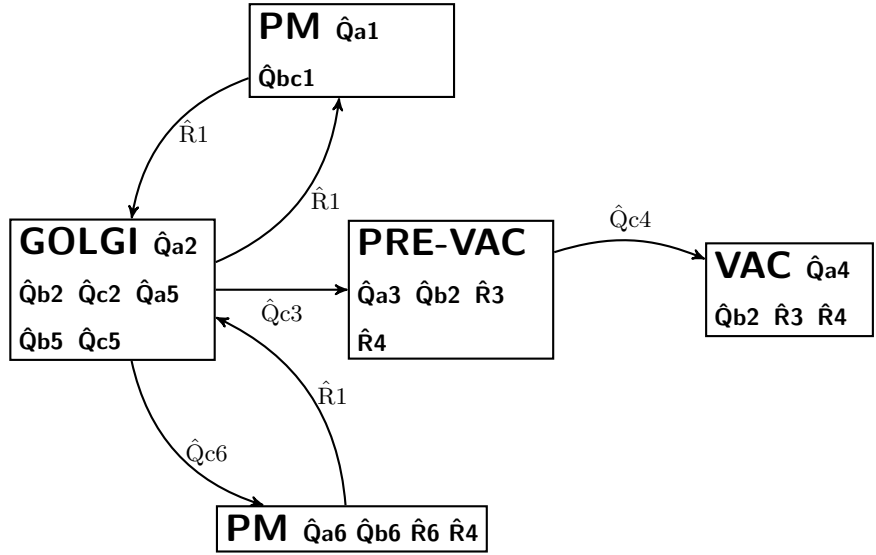


In figure~\ref{fig:yeast-vts}, we present the yeast VTS.
We have borrowed the VTS from~\cite{burri2004complete}.
It has been adapted from the paper by
separating the v and the t SNAREs. 
It is clear that it is an incomplete description of the VTS.
For example, the inactive molecules were not reported in the reference.
We are currently searching for more literature that can help us complete
all known information about the VTS.
%



\end{document}